\documentclass[epj]{svjour1}
\usepackage{latexsym}
\usepackage[english]{babel}
\usepackage{graphics}
 \usepackage{amssymb}

\journalname{Applied Physics B}

\begin{document}

\title{Hexagonal microlasers based on organic dyes in nanoporous crystals}

\author{Ingo Braun\inst{2} %
 \and Guido  Ihlein\inst{1} \and Franco Laeri\inst{3} %
 \and Jens U. N{\"o}ckel\inst{4} \and G{\"u}nter Schulz-Ekloff\inst{2} %
 \and Ferdi Sch{\"u}th\inst{1} \and Uwe Vietze\inst{3}%
 \and {\"O}zlem Wei{\ss}\inst{1} \and Dieter W{\"o}hrle\inst{2}
%
}                     
%
%
\mail{Franco Laeri, Institute of Applied Physics, Darmstadt
University of Technology, D-64289 Darmstadt-Germany; Email:
franco.laeri@physik.tu-darmstadt.de,  Tel.: +49-(0)6151-165495,
Fax: +49-(0)6151-163022}
 \institute{Max-Planck-Institut f{\"u}r Kohlenforschung,
 D-45470 M{\"u}lheim, Germany, Fax: +49-(0)208-3062995
 \and University of Bremen, D-28334  Bremen, Germany, Fax: +49-(0)421-2184935
 \and Darmstadt University of Technology, D-64289 Darmstadt,
 Germany, Fax: +49-(0)6151-163022
 \and  Max-Planck-Institut f{\"u}r Physik komplexer
 Systeme, D-01187 Dresden, Germany, Fax: +49-(0)351-8711999}
\date{23 February 2000}
\abstract{Molecular sieves, such as nanoporous AlPO$_4$-5, can host
a wide variety of laser active dyes. We embedded pyridine 2
molecules as a representative of a commercially available dye which
fits into the channel pores of the host matrix. Many efficient dye
molecules, such as rhodamines, do not fit into the pores. But
modifying the structure of the dyes to appear like the used
templates allows to increase the amount of encapsulated dyes. The
properties of resulting microlasers depend on size and shape of the
microresonators, and we discuss a model for microscopic hexagonal
ring resonators. In terms of pump needed to reach lasing threshold
molecular sieve microlasers are comparable to VCSELs. For dyes
which fit into the pores we observed a partial regeneration of
photo-induced damage.
 }

\maketitle

\section{Introduction}
\label{intro}

In recent years dye laser emission from from crystalline materials
attracted considerable interest \cite{RIF95,KAH96}. At the same
time in solid state dye lasers based on polymer matrices single
mode operation at high peak powers was demonstrated \cite{DUA95},
and photodegradation processes were identified \cite{POP99}.
Recently, we reported on microscopic dye lasers, in which the
active molecules were embedded in crystallographically defined
nanometer size pores of molecular sieves \cite{VIE98}. In the past,
such nanoporous materials, e.g.\ zeolites, proved to be pertinent
especially to the catalysis in oil refining and petrochemistry,
where they are used in huge amounts. It is only recently that
applications of this class of porous crystalline material are
discussed in an optical context. In fact, molecular sieves can be
used to host a wide variety of optically relevant guests, such as
atoms, ions, or molecules \cite{STU90}. Sieves with tubular or
channel pores, in particular, can act as an ordering framework for
guest molecules \cite{SUL94}. For optical applications it is
possible to exploit thus distinct spatial symmetries in the
arrangement of organic molecules, enhancing so, for example, the
second order nonlinear susceptibility of the compound \cite{COX88},
or admitting anisotropic F{\"o}rster energy migration \cite{GFE98}.
Nanoporous crystals were also studied as hosts for luminophores
\cite{BRE91}, pigments \cite{BRA97}, optical switches \cite{HOF97}
or dye lasers \cite{VIE98}. In the following we report about the
novel emission properties of hexagonal monolithic microresonators
realized with dyes intercalated in molecular sieve microcrystals,
and the photostability of these systems.

\begin{table}
\caption{Lattice constants and free pore diameter $\phi$ of
molecular sieves with linear channels \protect\cite{MEI96}.}
\label{tab_pores}
\begin{center}
\begin{tabular}{lcccc}
 \hline\noalign{\smallskip}  & $a$/nm & $b$/nm & $c$/nm & $\phi$/nm  \\
 \noalign{\smallskip}\hline\noalign{\smallskip}
  & \multicolumn{4}{c}{hexagonal}\\
 mazzite & 1.84 & & 0.76 &0.74 \\
 AlPO$_4$-5 & 1.34 && 0.84 & 0.73 \\
 zeolite L & 1.84 && 0.75 & 0.71 \\
 gmelinite & 1.38 && 1 & 0.7 \\
 offretite & 1.33 && 0.76 & 0.68 \\
 CoAPO-50 & 1.28 && 0.9 & 0.61 \\
 cancrinite & 1.28 && 0.51 & 0.59 \\
  & \multicolumn{4}{c}{orthorhombic}\\
 AlPO$_4$-11 & 1.35 & 1.85 & 0.84 & 0.63$\times$0.39\\
 mordenite & 1.81 & 2.05 & 0.75 & 0.70$\times$0.65\\
   \noalign{\smallskip}\hline
\end{tabular}
\end{center}
\end{table}

\section{Material}
\label{material}

\subsection{Host crystal}
\label{host}

Molecular sieve materials are characterized by a
crystallographically defined framework of regularly arranged pores.
In table \ref{tab_pores} we list some sieves, which, owing to their
large diameters of channel pores, are suitable for hosting
optically effectual organic molecules. Among the listed materials
especially the aluminophosphate AlPO$_4$-5 (molecular mass
1463.4~g/mol) can be synthesized with good optical transparency and
low internal scattering losses. Its channel pores exhibit a
diameter of 0.73~nm, which is large enough to accomodate suitable
organic dye molecules.

AlPO$_4$-5--crystals are crystallized from aqueous or alcoholic
solutions under hydrothermal conditions, with the addition of an
organic structurizing agent, called template. The template,
tri-n-propylamine in our case, is necessary to direct the synthesis
towards the desired structure. The preferred pH range for the
synthesis is mildly acidic to mildly basic. The most utilized
source of phosphore is orthophosphoric acid, and the most studied
sources of aluminum are pseudoboeh\-mite and alkoxides
\cite{WIL91}. Single crystals with near\-ly perfect morphology were
grown using hydrofluoric acid \cite{GIR95}. With a specially
prepared and aged aluminum hydroxide gel crystal sizes around  1~mm
in $c$-axis direction were obtained \cite{SUN96}, whereas microwave
heating proved to increase the crystallization rates by more than
one order of magnitude \cite{GIR95,DU97}.

\begin{figure}[tbh]
\resizebox{0.48\textwidth}{!}{%
  \includegraphics{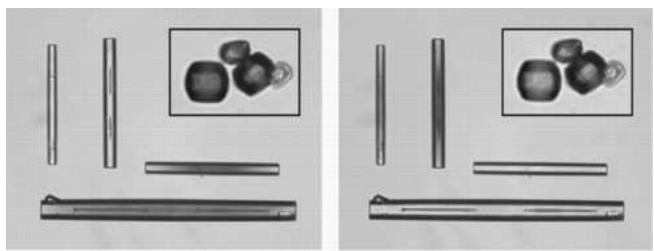}
}
 \caption{Transmission micrographs of the dichroism in dye-loaded
AlPO$_4$-5 crystals; the rod-shaped crystals contain ca.~0.1~wt-\%
or 260 pyridine~2 molecules per unit cell, while the barrel-shaped
ones (shown in the inset) enclose rhodamine~BE50 (ca.~0.5~wt-\% or
75 molecules per unit cell). Only the polarization component
parallel to the optical transition moment of the molecules is
absorbed. In the rod-shaped crystal the pyridine~2 dyes are
completely aligned, whereas with the rhodamine~BE50 dye in the
barrel-shaped crystals we observe only a weak dependence of the
color upon the incident polarization. {\bf Left:} incident light
horizontally polarized; {\bf Right:} incident light vertically
polarized.}
 \label{dichro}
\end{figure}

Pure AlPO$_4$-5 crystals are optically transparent from below
400~nm to above 800~nm ($n_{500\,{\rm nm}}=1.466$), and after
removing the template (usually by heating/calcinating) they exhibit
practically no birefringent properties. X-ray patterns revealed
systematic absences which are consistent with space group $\rm P
\frac{6}{\rm m}cc$ as well as P6cc, whereby, however, the latter
implies a polar nature of AlPO$_4$-5 single crystals: In fact, the
4 in the formula is the result of the strict alternation of Al and
P in the tetrahedral nodes of the framework, which prevents the
corner-sharing oxygen tetrahedra to occur with odd numbers, and
which leads to an alternating stacking of Al and P in the direction
of the channels ($c$-axis). This is assumed to cause the
crystallographic polar nature \cite{BEN83} of the framework
\cite{MAR94}. The macroscopic polar nature of AlPO$_4$-5 single
crystals has been proven recently in scanning pyroelectric
microscopy investigations \cite{KLA99}. There it was also observed
that AlPO$_4$-5 crystals are usually twinned. The murky stripes
inside the pyridine~2-loaded crystals shown in Fig.~\ref{dichro}
and their slightly bowed side faces could well be a result of this
kind of twinning. It is not clear, however, to what extent such
twinning should affect the optical properties relevant for the
effects discussed here.

\begin{figure}
\begin{center}
\resizebox{0.4\textwidth}{!}{%
  \includegraphics{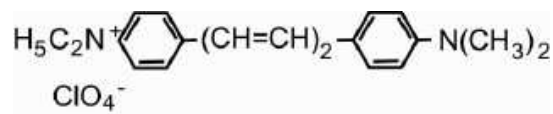}
}
 \caption{Structure formula of the dye 1-ethyl-4-(4-({\sl
p}-dimethylamino\-phen\-yl)-1,3-butadienyl)-pyridinium perchlorate
(pyr\-i\-dine~2 \protect\cite{BRA94}); molecular mass 378.9~g/mol.}
 \label{pyrid2}
\end{center}
\end{figure}

\begin{figure}
\resizebox{0.48\textwidth}{!}{%
  \includegraphics{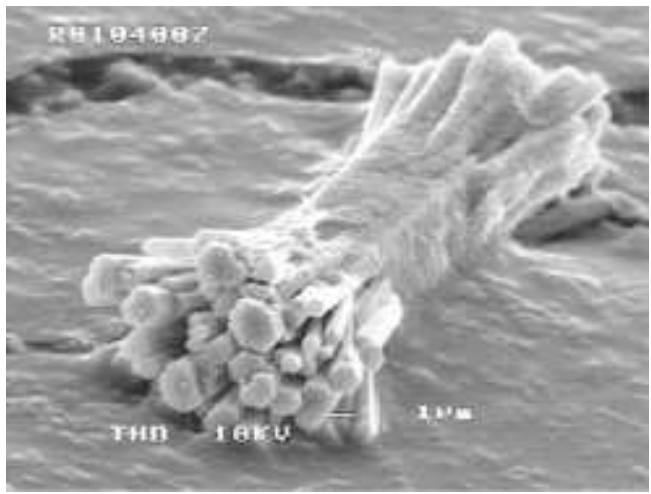}
 }
\caption{Morphology of typical pyridine 2/AlPO$_4$-5 crystals with
lasing properties.}
 \label{fasc}
\end{figure}

\subsection{AlPO$_4$-5/pyridine~2 compound}
\label{pyrid}

The compounds were synthesized following a procedure in which the
dye 1-ethyl-4-(4-({\sl
p}-dimethylamino\-phen\-yl)-1,3-butadienyl)-pyridinium perchlorate
(cf.\ Fig.~\ref{pyrid2}) (pyr\-i\-dine~2 \cite{BRA94}) is added to
the template or to the aluminum hydroxide suspension
\cite{HOP93,DEM95,IHL98}. After 1~h of hydrothermal synthesis dark
red crystals with a length of up to 100~$\mu$m were obtained. The
crystals were refluxed in ethanol for 4~h whithout any detectable
extraction of dye. The slim, linear dye molecules fit snugly into
the 0.73~nm wide channel pores of the nanoporous AlPO$_4$-5 host,
resulting so in an ordered compound material, in which the dye
molecules are all aligned along the crystal $c$-axis. As a
consequence the compound exhibits strong dichroism, and the emitted
fluorescence light is polarized parallel to the $c$-axis. This is
documented in Fig.~\ref{dichro}. It is also observed that with the
inclusion of pyridine~2 the entire compound acquired pyroelectric
properties and an optical second order susceptibility \cite{VIE98}.

Depending on the dye content different morphologies of crystals are
observed. Regular hexagonal crystals with a rodlike form, as e.g.\
the ones shown in Fig.~\ref{dichro}, were obtained when the dye
content was low, around 0.1~wt\% or 260 molecules per unit cell. At
higher concentrations the dye accumulated in the middle of the
crystal, and the morphology was severely disturbed. At dye content
$\gtrapprox$~0.2~wt\% crystals with a caracteristic fascicular
shape resulted; cf.\ Fig~\ref{fasc}. Given the small size of the
crystals we had to determine the dye content by chemically
dissolving them. With this method, however, it was not possible to
accurately determine the spatial distribution of the dye. We
therefore evaluated the content qualitatively by comparing the
depth of the color. It was only with these fascicled
pyridine~2-loaded crystals that we observed laser emission.
Apparently the low concentration of dye in the undisturbed
rod-shaped crystals did not spoil the growth but on the one hand
was not sufficient to provide the necessary optical gain for
compensating all losses.

\begin{figure}
\begin{center}
\resizebox{0.4\textwidth}{!}{%
  \includegraphics{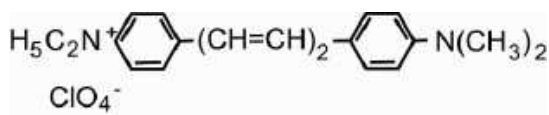}
}
 \caption{Structure formula of the new dye rhodamine BE50 (ethanaminium,
$N$-[6-(diethylamino)-9-[2-($N,N$-di\-meth\-yl-3-amino-1-%
propoxycarbonyl)phenyl]-3H-xanthen-3-y\-li\-dene]-$N$-ethyl-chloride)
molecular mass 564~g/mol \protect\cite{BOC98}. }
 \label{rhbe50}
\end{center}
\end{figure}

\subsection{AlPO$_4$-5/rhodamine~BE50 compound}
\label{rhod}

The synthesis gel was prepared according to recipes
\cite{BEN83,WIL82} modified for the purpose of crystallization
inclusion of dyes \cite{WOH92}. To a suspension of 61.6~mmol $\rm
Al_2O_3$ (8.44~g; Pural~SB, Condea Chemie) as aluminum source and
75~g deionized water, 61.6~mmol $\rm P_2O_5$ (14.20~g phosphoric
acid; 85~wt\%, p.a.\ Merck) in 11.3~g deionized water was added
under mechanical stirring. After 5~min a uniform gel formed and
then 92.4~mmol tripropylamine (13.25~g $\rm Prop_3N$, Merck) was
added slowly. Subsequently, the appropriate amount (0.1--10~mmol)
of the dye powder was mixed with the gel. The new derivative
rhodamine~BE50  (ethanaminium,
$N$-[6-(diethylamino)-9-[2-($N,N$-di\-meth\-yl-3-amino-1-%
propoxycarbonyl)phenyl]-3H-xanthen-3-y\-li\-dene]-$N$-ethyl-chloride;
cf.\ Fig.~\ref{rhbe50}) was synthesized by esterification of
rhodamine~B (Rh~B) with 3-dimethyla\-mi\-no-1-propanol
\cite{BOC98}. It was shown that the concentration of rhodamine BE50
(Rh~BE50) achievable by crystallization inclusion in AlPO$_4$-5
exceeds the possible Rh~B concentration by a factor of 3--4. This
was attributed to the different molecule structures, i.e., to the
zwitterionic nature of Rh~B on one hand, and to the additional
positive charge of a protonated aliphatic amino group of Rh~BE50 on
the other hand \cite{BOC98}. The latter molecules with the
localized positive charge are more compatible with the AlPO$_4$-5
framework than the delocalized charge of Rh~B. As a consequence we
observe that at the same dye concentration  Rh~BE50 inclusion leads
to a better crystal morphology than Rh~B inclusion \cite{BRA99}.
The synthesis of the Rh~BE50/AlPO$_4$-5 crystals was performed by
microwave heating \cite{BRA98}, which has proven to be superior in
respect to avoiding damage of sensitive dyes like coumarines
\cite{BRA97} as well as in reducing the time of synthesis
\cite{BOC98}.

\begin{figure}
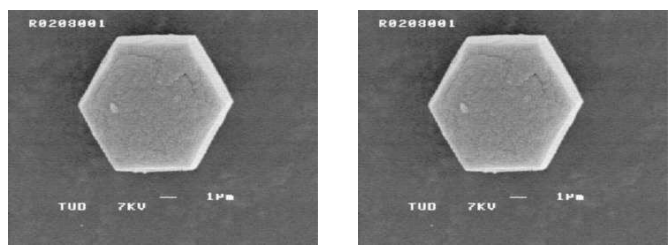

 \resizebox{0.23\textwidth}{!}{%
  \includegraphics{figures/BE50morphS.epsf}
 }
 \hfill
 \resizebox{0.23\textwidth}{!}{%
  \includegraphics{figures/BE50morphO.epsf}
 }
 \caption{Morphology of typical rhodamine BE50/AlPO$_4$-5 crystals
  with lasing properties; dye content ca.~0.5~wt-\% or 75 dye molecules
per unit cell.}
 \label{BE50morph}
\end{figure}

Unlike the pyridine~2 molecules, which with a diameter of  0.6~nm
fit into the 0.73~nm wide pores of the AlPO$_4$-5 host, the Rh~BE50
molecules with dimension of 0.91~$\times$~1.36~nm$^2$ must be
accomodated in defect sites (mesopores) of the host crystal.
Remarkably, up to concentrations of 1 molecule per 75 unit cells
this remains without any visible negative consequences for the
crystal morphology, as is documented in Fig.~\ref{BE50morph}.

In Fig.~\ref{dichro} the dichroic properties of the compound are
illustrated. In comparison with the pyridine-2/AlPO$_4$-5 compound,
the dichroism of the Rh~BE50 com\-pound is reduced and the
fluorescence emission is partially polarized. Thus it is inferred
that the anisotropy of the host structure does not fully carry
through the mesopores, and thus the guest molecules are only weakly
aligned.

\begin{figure}
\begin{center}
\resizebox{0.23\textwidth}{!}{%
  \includegraphics{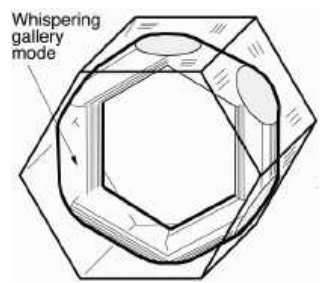}
 }
 \end{center}
\caption{Na\"ive ray picture interpretation of a whispering gallery
laser mode inside a hexagonal prism. In the following it is shown
that for prisms with sizes of a few wavelengths this picture has to
be modified.}
 \label{whisp}
\end{figure}

\section{Microresonator structure}
\label{reson}

As is visible in Fig.~\ref{dichro} -- and as the polarization of
the emitted fluorescence of the compounds indicates -- the
absorption, as well as the emission dipole moment of the included
dyes are oriented preferentially along the crystal $c$-axis
(Complete orientation in the pyridine 2/AlPO$_4$-5 compound). As
dipole emission along the dipole-axis is not possible, the emission
parallel to a plane perpendicular to the prevailing dipole
orientation (i.e.\ the hexagonal axis) is enhanced. Here a bundle
of emission directions meets the condition for total internal
reflection (TIR) at the hexagonal side faces inside the crystal. In
a whispering-gallery-mode-like way the corresponding emission can
circulate sufficiently often to accumulate the gain required to
overcome the lasing threshold; cf.\ Fig.~\ref{whisp} in which this
intuitive model is illustrated for a particular ray bundle of high
symmetry. However, with resonator sizes in the order of a few
wavelengths, as discussed here, the ray picture does not represent
the field modes even in a qualitatively correct way. E.g.\ the ray
picture insinuates a mode concentration in the center of the faces
and field-free corners. However, this is not consistent with the
experimental evidence, which clearly shows that the emission occurs
at the corners; cf.\ Fig.~\ref{nearfield}. In fact, for dielectric
resonator structures a few wavelengths large with TIR field
confinement a wave model has to be worked out.

\subsection{Ray picture}

The main feature that distinguishes the hexagonal resonator from
other common whispering-gallery type cavities such as microdroplets
\cite{mekis95} or semiconductor disk lasers \cite{nature,gmachl},
is that the latter do not exhibit sharp corners and flat sides.
Portions of the boundary in convex resonators can act as focussing
or defocussing elements, but the straight sides of a hexagon are
neither one nor the other. The hexagon in fact constitutes a
self-assembled realization of a {\em pseudointegrable} resonator
\cite{richensberry}: For a polygon with precisely $120^{\circ}$
angles between adjacent sides, any ray lauched at some angle to the
surface will go through only a finite number of different
orientations \cite{hobson}, just as in the more familiar
rectangular resonator where there are at most two non-parallel
orientations for any ray path. In the hexagon, a ray encounters the
interface with at most three different angles of incidence. Despite
this apparent simplicity, there exists no orthogonal coordinate
system in which the wave equation for the hexagonal cavity can be
solved by separation of variables. This property of
non-integrability is shared by wave equations whose classical
(short-wavelength) limit exhibits chaos. However, ray paths in the
hexagon display a degree of complexity that cannot be classified as
chaotic, and hence the term pseudointegrability has been coined for
these systems. The ray-wave duality in ``billiards'' of this type
has to be addressed in order to explain how they can support
whispering-gallery modes that emit at the corners.

\begin{figure}[bt]
 \begin{center}
\resizebox{0.48\textwidth}{!}{%
  \includegraphics{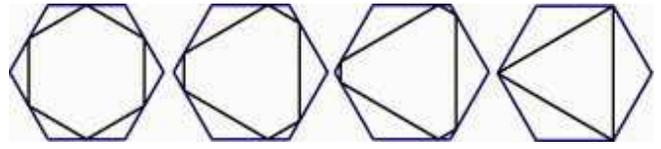}
 }
 \end{center}
\caption{\label{fig:hexaorbits}
 All four orbits shown have the same length and the same
angle of incidence with respect to the interface normal,
$\chi=60^{\circ}$. They hence satisfy the condition for total
internal reflection, $\sin\chi>1/n=0.69$. In the case of rays
impinging on the corners (rightmost picture) the ray picture breaks
down.}
\end{figure}

In order to characterize the sample size, one can specify either
the radius $R$ of the hexagon at the corner points or -- more
conveniently -- the width over flats (WoF) satisfying the relation
$R= WoF / \sqrt{3}$. The closed ray path underlying
Fig.~\ref{whisp} is only one member of an infinite family of
periodic orbits of the hexagon billiard that all have the same
length, $L=3\times WoF$, as shown in Fig.~\ref{fig:hexaorbits}.
Long-lived cavity modes should be expected only if the
corresponding rays satisfy the condition of TIR at the interface,
$\sin\chi>1/n$, where $\chi$ is the angle of incidence with respect
to the surface normal. This is true for the orbits of
Fig.~\ref{fig:hexaorbits}. In a naive ray approach one would
furthermore obtain the spectrum of modes by requiring an integer
number of half wavelengths to fit into $L$, leading to constructive
interference on a round-trip. As we shall see shortly, this
estimate is justified, even though a proper treatment of the
ray-wave connection has to take into account that any given mode is
in fact made up of a whole {\em family} of different ray paths. The
ultimate breakdown of the ray model, however, is illustrated in
Fig.~\ref{fig:hexaorbits} by the degenerate ray orbit hitting the
corners where the classical laws of refraction and reflection
become undefined.

\subsection{Wave picture: spectral properties}\label{WavPic}

\begin{figure}[bt]
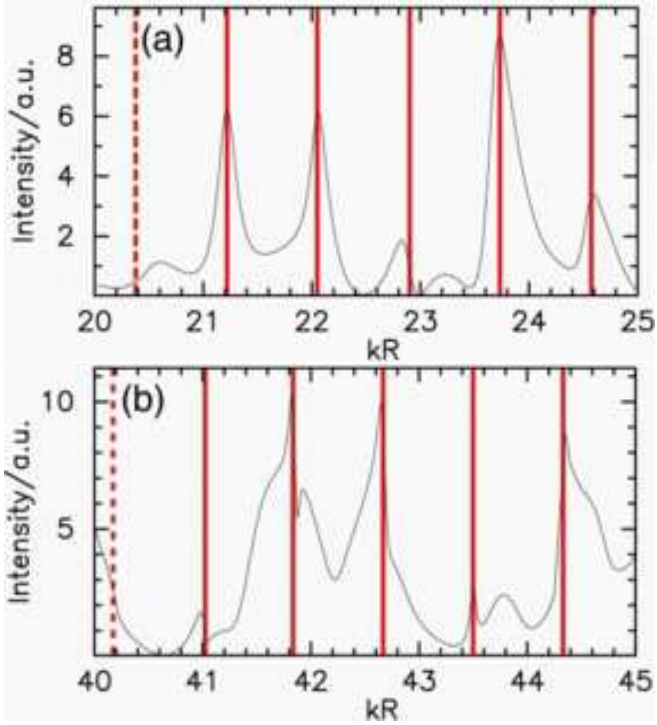

\begin{center}
\resizebox{0.48\textwidth}{!}{%
  \includegraphics{figures/spec1b.epsf}}
  \vspace*{5mm}
\resizebox{0.48\textwidth}{!}{%
  \includegraphics{figures/spec2b.epsf}}
\end{center}
\caption{\label{fig:k20spec} Calculated scattering intensity
spectra of a hexagonal cylinder for plane-wave incidence at
$15^\circ$ to a side face and detection at $60^{\circ}$ from
incidence. (a) corresponds to a spectral interval $\lambda\approx
653\ldots 816$~nm for width over flats (WoF) $4.5\,\mu$m; (b)
covers the interval $\lambda\approx 605\ldots 680$ nm for WoF
$7.5\,\mu$m. Vertical lines are guides to the eye, indicating
narrow resonances. The spacing between resonances is
$\Delta(kR)\approx0.84$ in (a) and $\Delta(kR)\approx0.83$ in (b),
in good agreement with the characteristic mode spacing
$\Delta(kR)_c\approx0.83$ of a closed hexagonal orbit. Expected
resonances not clearly seen in the above spectra are marked by
dashed lines; they appear at other detection angles.}
\end{figure}

Because the hexagonal faces are neither focussing nor defocussing,
there is no obvious way of determining the weight that should be
given to individual members of a ray family as depicted in
Fig.~\ref{fig:hexaorbits}, in order to predict the spatial
structure of the resulting mode. Full solutions of Maxwell's
equations have therefore been carried out for the TM polarized
modes of a dielectric hexagonal prism, using methods previously
applied in \cite{nature,gmachl}. In anticipation of the
experimental spectra to be discussed in section 4, attention here
is focused on three different sample sizes, with WoF $4.5\,\mu$m,
$7.5\,\mu$m and $22\,\mu$m. The aim is to understand the observed
laser line spacings and the emission directionality. Comparison of
the calculated and observed linewidths will not be attempted
because the simulations do not take gain-narrowing into account.

In order first to verify that orbits of the type shown in
Fig.~\ref{fig:hexaorbits} determine the characteristic mode spacing
of these cavities, Fig.\ \ref{fig:k20spec} shows light scattering
spectra for different sample sizes in the vicinity of the
experimental wavelengths. Intensity is plotted versus dimensionless
wavenumber $kR$, where $k=2\pi/\lambda$. This is the natural scale
for comparison with semiclassical predictions because modes
differing by one node along a closed path should then be equally
spaced, with a characteristic separation
$\Delta(kR)_c=2\pi\,R/(n\,L)=2\pi/(3\sqrt{3}n)=0.825$ independent
of the sample size. The expected wavelength spacing of the modes
(free spectral range $FSR$) is $\Delta\lambda=\lambda^2 \times
\Delta(kR)_c / (2\pi R) \approx 23$~nm in (a) and
$\Delta\lambda\approx 11$~nm in (b). For $WoF = 22\,\mu$m, we
obtain $\Delta\lambda\approx 4.9$~nm. Figure \ref{fig:k20spec}
indeed shows a series of resonant features with approximately the
predicted wavevector spacing.

Each of the peaks marked in Fig.~\ref{fig:k20spec} (b) is in fact a
multiplet, which is not resolved because the splittings of the
individual modes comprising the multiplet are smaller than their
passive linewidths. There is evidence for this because several of
the peaks are very asymmetric and in particular exhibit a steep
slope on one side. For an isolated resonance, the most general
lineshape that could arise is the Fano function (of which the
Lorentzian is a special case), which however does not yield
satisfactory fits here.

\begin{figure}[bt]
\begin{center}
\resizebox{0.48\textwidth}{!}{%
  \includegraphics{figures/spec3.epsf}}
\vspace*{5mm}
\resizebox{0.48\textwidth}{!}{%
  \includegraphics{figures/spec4.epsf}}
\end{center}
\caption{\label{fig:k40smoothspec} Calculated scattering intensity
spectra for slightly rounded hexagonal cavities (shapes depicted as
insets). The incoming plane wave is at an angle of $15^{\circ}$ to
a facet in (a) and $30^{\circ}$ in (b); detection occurs at
$60^{\circ}$ from incidence. Spacings between modes of the same
color agree well with $\Delta(kR)_c\approx 0.83$, cf.\
Fig.~\ref{fig:k20spec}. All resonances in (a) appear as doublets.
At $kR\approx 42.5$ the doublet structure is seen most clearly. In
(b), stronger deviation from hexagonal shape leads to further
lifting of degeneracies. Dashed lines mark expected resonances not
seen at this observation angle.}
\end{figure}

\begin{figure}
\begin{center}
\resizebox{0.48\textwidth}{!}{%
  \includegraphics{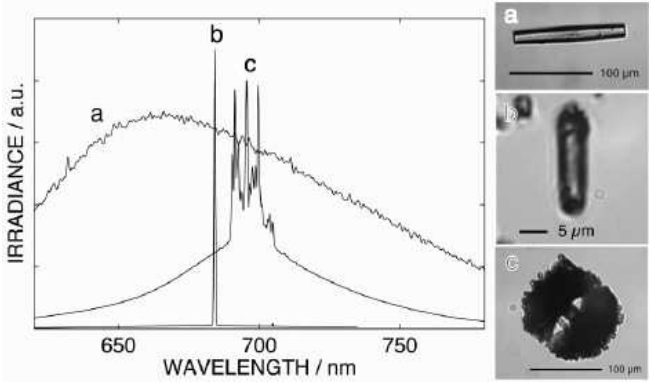}
 }
 \end{center}
\caption{Emission and lasing spectra of different pyridine~2-loaded
compounds with from {\bf a} to {\bf c} increasing amounts of
included dye. The dye concentration was estimated empirically based
on the sample color depth. The width over flats of sample {\bf b}
is 4.5~$\mu$m. The free spectral range of this resonator is so
large (ca.~24~nm) that one emission mode acquires the available
gain resulting in single line emission.
 }
 \label{pyr2spectr}
\end{figure}

To further expose the multiplet structure, we modeled deviations
from the ideal hexagonal shape which could lead to narrower
individual linewidths and increase the multiplet splitting. Shape
perturbations were chosen that preserve the $D_{6h}$ point group
symmetry and hence remove only ``accidental'' quasi-degeneracies.
The actual perturbation that is present in the samples of
Figs.~\ref{pyr2spectr} and \ref{BE50spectr} eluded experimental
characterization, so that a model calculation can reproduce only
generic features which are insensitive to the precise type of
perturbation. One such feature is the {\em average} mode spacing
after degeneracies have been lifted sufficiently.

Figure \ref{fig:k40smoothspec} (a) shows the spectrum of a rounded
hexagon where the radius of curvature at the corners is $\rho
\approx 0.9\,\lambda$ (assuming $\lambda\approx 610$ nm for
definiteness). No qualitative difference to Fig.~\ref{fig:k20spec}
(b) is seen, except that the resonant features have become somewhat
narrower, thus enabling us to identify two distinct series of modes
with caracteristic spacing $\Delta(kR)_c$. This indicates that
departures from sharp corners are not resolved in the wave equation
when their scale is smaller than $\lambda$. A qualitatively
different spectrum is observed in Fig.~\ref{fig:k40smoothspec} (b)
where $\rho\approx 3.7\,\lambda$. Here, the perturbation reveals
three well-separated, interpenetrating combs of modes, again with
period $\Delta(kR)_c$. There are $21$ distinct resonances in the
wavelength interval of Fig.~\ref{fig:k40smoothspec} (b), which
translates to an average mode spacing of $\Delta\lambda\approx 3.6$
nm for a $WoF = 7.5\;\mu$m resonator, well in agreement with the
experiment; cf.\ Fig.~\ref{BE50spectr}.

In order to verify that no further modes will be revealed by other
choices of deformation, an independent estimate of the average
density of modes can be made based on semiclassical considerations
\cite{noeckelunpublished}:

\begin{eqnarray}
\left\langle\frac{dN}{d(kR)}\right\rangle&=&\frac{n^2\,k\,R}{4}\nonumber\\
&\times&\left[1-\frac{2}{\pi}\,\left(\arcsin{
\frac{1}{n}}+\frac{1}{n}\, \sqrt{1-\frac{1}{n^2}}\right)\right]
\label{wgweylformtseqn}
\end{eqnarray}

Here, $dN$ is the number of modes in the interval $d(kR)$. The
result is $\langle\frac{dN}{d(kR)}\rangle\approx 4.6$, and hence we
expect $\approx 22$ modes in the interval of
Fig.~\ref{fig:k40smoothspec} (b), again in good agreement with the
actual count.

\subsection{Wave picture: intensity profile}

There is one class of quasi-degeneracies that is not removed by any
of the perturbations in Fig.~\ref{fig:k40smoothspec}: their
physical origin is time-reversal symmetry for the ray motion inside
the cavity. Any of the periodic orbits in Fig.~\ref{fig:hexaorbits}
can be traversed either clockwise or counterclockwise, and the same
holds for more general ray paths. The different propagation
directions can be linearly combined in various ways to obtain
nearly-degenerate standing-wave patterns that differ only in their
parity with respect to some of the crystal's reflection axes. A
minute splitting does exist because the non-integrability of the
ray motion implies that the propagation direction itself is not a
``good quantum number'', i.e. reversals of the sense of rotation
are unlikely but not impossible in the wave equation. This is
analogous to quantum tunneling and hence leads only to
exponentially small splittings that can be neglected on the scale
of the individual resonance linewidths \cite{DAV81}. These
multiplets have been counted as one resonance in Eq.\
(\ref{wgweylformtseqn}).

\begin{figure}[hbt]
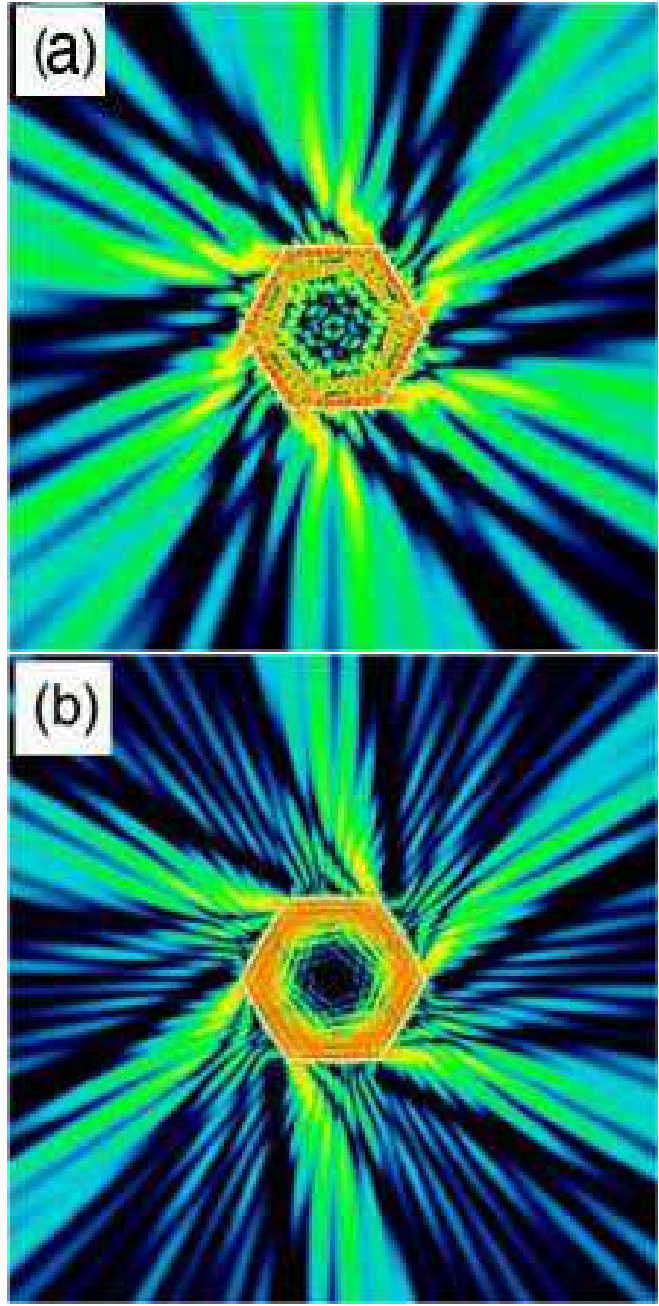

\begin{center}
\resizebox{0.48\textwidth}{!}{%
  \includegraphics{figures/wavesharpa.epsf}} \vspace*{5mm}
\resizebox{0.48\textwidth}{!}{%
  \includegraphics{figures/wavesharpb.epsf}}
\end{center}
\caption{\label{fig:waksharp} False-color representation of the
cross-sectional intensity in the ideal hexagon for a mode with (a)
$kR=22.89$ [cf.\ Fig.~\ref{fig:k20spec} (a)] and (b) $kR= 42.78$
[cf.\ Fig.~\ref{fig:k20spec} (b)]. The resonance width is
$\delta(kR)=0.10$ in (a) and $\delta(kR)=0.04$ in (b).}
\end{figure}

Following this reasoning, in Fig.\ \ref{fig:waksharp} the {\em
traveling-wave} patterns belonging to one of the resonances in
Fig.\ \ref{fig:k20spec} (a) and (b), respectively, is plotted.
High-intensity ridges inside the resonator form a
whispering-gallery-like pattern that decays from the interface into
the cavity center. The number of ridges in the radial direction
(perpendicular to a side face) provides an approximate analogue of
a transverse mode order, however upon closer examination one sees
that the number of ridges and nodal lines is not uniquely defined,
in particular along a diameter joining opposite corners. The modes
can therefore not be properly labeled by ``good quantum numbers''
characterizing the number of radial and azimuthal nodes -- this is
a direct consequence of the nonintegrability of the problem. The
most significant difference to the whispering-gallery modes of a
circular cavity is clearly the anisotropic emission. High intensity
is seen to emanate predominantly from the corners and is directed
almost parallel to an adjacent crystal facet. The overall emission
pattern is very similar in both modes despite the large difference
in size (or $kR$) between the two hexagons.

\section{Laser properties}
\label{laserprop}

The microcrystals were pumped with 10~ns pulses from the 532~nm
second harmonic of a Nd:YAG-laser delivered to the sample with an
optical fiber. The emitted luminescence was collected with a
$20\times$-microscope objective (collecting aperture 42$^\circ$)
relaying the microlaser emission simultaneously to the spectrometer
(cooled CCD-detector Oriel InstaSpecIV), and to the imaging system
consisting of a cooled low noise CCD-camera (PCO SensiCam). In
Fig.~\ref{pyr2spectr} the emission and lasing spectra of different
pyridine~2-loaded compounds are shown. With increasing dye content
the fluorescence emission maximum of regularly shaped crystals
shifted from 645~nm to 665~nm; cf.\ Fig.~\ref{pyr2spectr}a. In none
of these rod-shaped crystals laser emission was observed. However,
narrow laser emission peaks (observed linewidth of 0.3~nm given by
spectrometer resolution) occured in most fascicled samples.
Emission maxima in these fascicled samples were observed at
wavelenghts up to 695~nm, and again, increasing dye concentration
correlated with increasing redshift (cf.\ Figs.~\ref{pyr2spectr}b
and c). As already mentioned, together with increasing dye content
we also notice an increasingly disarranged crystal morphology. The
observed facts, disturbed morphology and red-shifted emission
spectrum are consistent with the hypothesis of a host-guest
interaction which increases with dye content. Essentially both, the
pyridine 2 molecules and the AlPO$_4$-5 framework carry a static
dipole moment \cite{BEN83,MAR94}. Obviously the buildup of
electrostatic energy in the crystal lattice has to be compensated
by an increasing amount of stacking faults. On the other side, the
mechanism of the redshift is not unequivocally identified, yet, but
is probably related to the one discussed in \cite{FOE49,FOE51}.

\begin{figure}
 \begin{center}
 \resizebox{0.35\textwidth}{!}{%
  \includegraphics{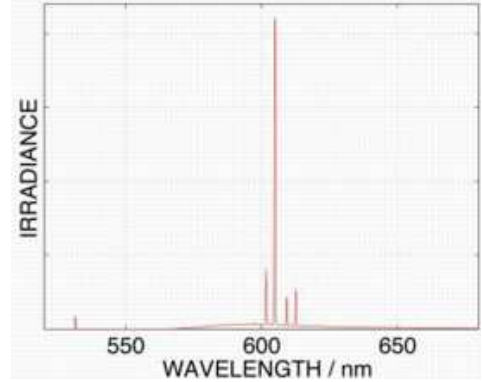}
 }
 \end{center}
\caption{Lasing spectrum of a rhodamine~BE50-loaded AlPO$_4$-5
microcrystal with a concentration around 75 unit cells per dye
molecule and size of 7.5~$\mu$m width over flats.}
 \label{BE50spectr}
\end{figure}

The dye concentration in the rhodamine~BE50-loaded AlPO$_4$-5
samples was around 75 unit cells per dye molecule, corresponding to
0.5~wt\%. Also with these samples the same correlation between the
emission wavelength and dye concentration was observed; cf.\ also
\cite{BOC98}. In contrast to the pyridine 2-loaded samples, the
fluorescence emission was not completely polarized. The observed
polarization contrast $c_p=
\frac{I_\parallel-I_\perp}{I_\parallel+I_\perp}$ was around 10\%,
indicating that in the average the Rh~BE50 molecules are only
weakly aligned parallel to the crystal $c$-axis.

\begin{figure}
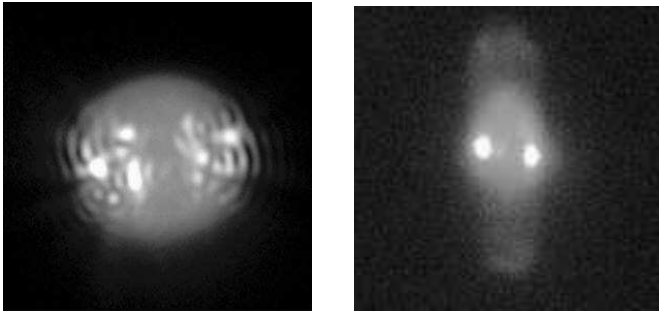

 \resizebox{0.23\textwidth}{!}{%
  \includegraphics{figures/BE50las.epsf}
 }
 \hfill
 \resizebox{0.23\textwidth}{!}{%
  \includegraphics{figures/pyr2las.epsf}
 }
 \caption{Patterns of the laser emission show that the emission
originates from regions along side edges. {\bf Left:}
rhodamine~BE50/AlPO$_4$-5 compound; width over flats 7.5~$\mu$m. An
electron micrograph of the sample is shown in
Fig.~\protect\ref{BE50morph} with horizontal $c$-axis. Here the
$c$-axis orientation is nearly vertical. The corresponding emission
spectrum is shown in Fig.~\protect\ref{BE50spectr}. {\bf Right:}
pyridine~2/AlPO$_4$-5 compound; width over flats 4.5~$\mu$m. The
corresponding sample and emission spectrum is represented in
Fig.~\protect\ref{pyr2spectr}~{\bf b}.
  }
 \label{nearfield}
\end{figure}

Independent of the type of loading, in most microcrystals with $WoF
\gtrapprox$~8~$\mu$m lasing was observed to occur on several sharp
lines with instrument resolution limited width. A typical example
is represented in Fig.~\ref{BE50spectr}. Note that the lines are
not equally spaced. In fact, the free spectral range ($FSR$) of
11~nm corresponding to the resonator size is far above the observed
line spacing of 3.2, 4.3 and 3.4~nm. This is in agreement with the
theoretical model of paragraph~\ref{WavPic}, in which the average
lasing mode spacing (after lifting the quasi-degeneracies in the
ideal hexagon) was estimated to be $\Delta\lambda =$ 3.6~nm. Also
in agreement with the theoretical discussion are the emission
regions where the laser light leaves the hexagonal resonators.
Figure~\ref{nearfield}(left) shows the laser emission as bright
spots. Clearly the emission is concentrated along the crystal
edges. The complex emission distribution is compatible with the
simultaneously recorded spectrum (cf.\ Fig.~\ref{BE50spectr}) which
reveals multimode emission.

\begin{figure}
\begin{center}
\resizebox{0.48\textwidth}{!}{%
  \includegraphics{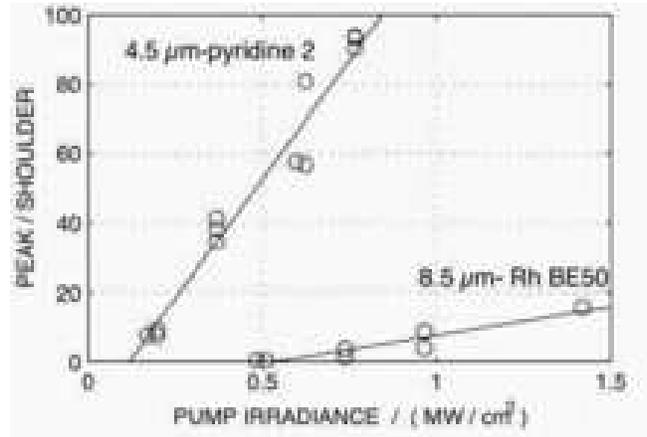}
 }
 \end{center}
\caption{Lasing threshold and differential efficiency of typical
AlPO$_4$-5/dye compounds. Shown is the peak of the laser emission
spectrum normalized by the fluorescence shoulder as a function of
the pump power density for the sample shown in
Figs.~\protect\ref{pyr2spectr}b and \protect\ref{BE50spectr}.
 }
 \label{threshold}
\end{figure}

On the other hand, samples with smaller resonator ($WoF
\gtrapprox$~4~$\mu$m), as e.g.\ the one shown in
Fig.~\ref{pyr2spectr}{\bf b} emitted one single laser line.
Therefore the emission is unadulterated by hole burning induced
multimode beating and interference, and appears in the simple
pattern shown in Fig~\ref{nearfield}(right), where two ca.\
1~$\mu$m-spots ($\approx$ microscope resolution limit) mark the
region of laser emission, which, again, is located on the crystal
edges. Compared with larger samples, the ratio of the line peak to
the underlying fluorescence shoulder of these small lasers is an
order of magnitude higher (cf.\ Fig.~\ref{threshold}).

While, as described above, the average linespacing of 3.6~nm
observed in the sample of Fig.~\ref{BE50spectr} is not compatible
with the free spectral range of $\Delta\lambda = 12$~nm of the
corresponding resonator, the 4.2~nm spacing of the 3 dominant peaks
in the sample shown in Fig.~\ref{pyr2spectr}c is in accord with the
$FSR$ resulting from the 22-$\mu$m-WoF hexagonal resonator. The
theoretical model for the $22\,\mu$m-WoF sample (cf.\
Eq.~(\ref{wgweylformtseqn})) yields an average mode spacing of
$\Delta\lambda\approx$ 0.5~nm which is close to the spectrometer
resolution. This high spectral density helps to explain the large
background in the lasing spectrum of Fig.~\ref{pyr2spectr}: It
appears likely that not all the individual lasing modes in this
sample were resolved, and hence part of the shoulder on which the
three peaks of curve c sit is probably made up of other lasing
modes. Any microcavity effects such as enhanced $\beta$ factor
(spontaneaous emission enhancement and suppression \cite{YAM96})
are also suppressed by the larger density of modes in the large
sample.

Figure~\ref{threshold} illustrates the differential efficiency
behavior of a typical microlaser with $WoF < 10\,\mu$m and one with
$WoF > 10\,\mu$m. Lasing threshold for the latter size samples was
around 0.5~MW/cm$^2$, regardless of the type of dye loading. On the
other side, crystals of smaller size ($WoF=4.5\,\mu$m) from the
same synthesis batch revealed a considerably smaller threshold
(0.12~MW/cm$^2$) and a factor of $>7$ larger differential gain.
Wether this is a consequence of quantum size effects \cite{YAM96}
will be clarified in progressing studies.

It is informative to compare the threshold of molecular sieve
microlasers with vertical cavity surface emitting lasers (VCSELs).
VCSELs with a size comparable to sample {\bf b} in
Fig.~\ref{pyr2spectr} exhibit threshold currents of ca.~1~mA, which
corresponds to $6.25 \times 10^{15}$~s$^{-1}$ electrons. On the
other hand, the threshold power density of 0.12~MW/cm$^2$ incident
on the molecular sieve laser surface of $1 \times 4.5\,\mu$m (cf.\
Fig.~\ref{nearfield}) corresponds to a current of $1.4 \times
10^{16}$~s$^{-1}$ 532-nm-photons, or a factor of 2.24 over the
electron rate. As the pump radiation is not polarized but the
molecular ensemble of the considered laser is aligned, only half of
the pump photons actually contribute to the inversion. Thus, in
terms of elementary (quantum) pump processes needed to reach lasing
threshold the molecular sieve lasers are comparable to VCSELs.

\section{Photostability}
\label{phstab}

\begin{figure}
 \begin{center}
\resizebox{0.48\textwidth}{!}{%
  \includegraphics{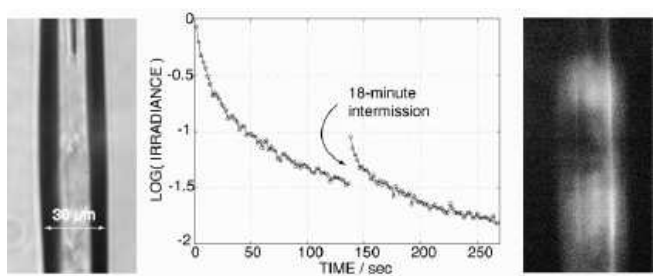}
 }
 \end{center}
\caption{Fluorescence activity of a pyridine~2-loaded AlPO$_4$-5
sample under bleaching laser irradiation. {\bf Left:} Micrograph of
the sample crystal. {\bf Center:} Fluorescence activity as a
function of time: After a first bleach period of 140~seconds the
bleach beam is interrupted for 18~minutes. During this intermission
the fluorescence recovers to start the second bleach period with
$3\times$ stronger emission. {\bf Right:} The bleaching laser is
incident from the left and concentrated in the center of the
crystal. Shown is the fluorescence distribution at the end of the
first bleach period of 140~s. Clearly visible is the bleached hole
in the center, where the bleach beam was concentrated.}
 \label{pyr2bleach}
\end{figure}

Photostability is a critical issue with dye lasers. We investigated
samples exhibiting an undisturbed morphology, similar to the one
shown in Fig.~\ref{dichro}. The pyridine 2-loaded samples were
irradiated with 10~Hz trains of 10~ns pulses of the 532~nm second
harmonic of a Nd:YAG-laser and a power density of 5~MW/cm$^2$.
Figure~\ref{pyr2bleach} illustrates the dwindle of fluorescence
activity of a pyridine~2-loaded AlPO$_4$-5 sample under such
bleaching irradiation. After a bleaching period of 140~seconds the
exposure was interrupted for 18~minutes. Then the bleaching
procedure was resumed. Apparently the fluorescence recovers during
the intermission. If it is assumed that bleaching consists in
breaking bonds of the dye molecules, then bond energies in the
eV-range have to be considered. Even if the dye debris might remain
encaged in their pores, spontaneous or thermally activated self
healing of broken eV-bonds seems not very probable. One can
therefore suppose that the recovery is due to diffusion of new,
intact dye molecules into the bleached volume. Considering the
stereometrically restricted possibilities inside the molecular
sieve framework together with diffusion distances of several $\mu$m
and the observed recovery time in the range of minutes, diffusion
is more plausible than self healing.

As bleaching reduces the concentration of dye, a blue\-shift of the
fluorescence is expected with increasing photobleaching
\cite{BEN83,MAR94}. However, the 656-nm-fluorescence emission
maximum of this sample is already at the shortest observed
wavelength (cf.\ Fig.~\ref{pyr2spectr}), corresponding to a low dye
concentration, and consequently, to weak dipole interactions. Thus,
the blueshift under these circumstances must be rather small. This
explains the fact that we could not detect a measurable blueshift
with these pyridine~2-loaded samples.

\begin{figure}
 \begin{center}
\resizebox{0.48\textwidth}{!}{%
  \includegraphics{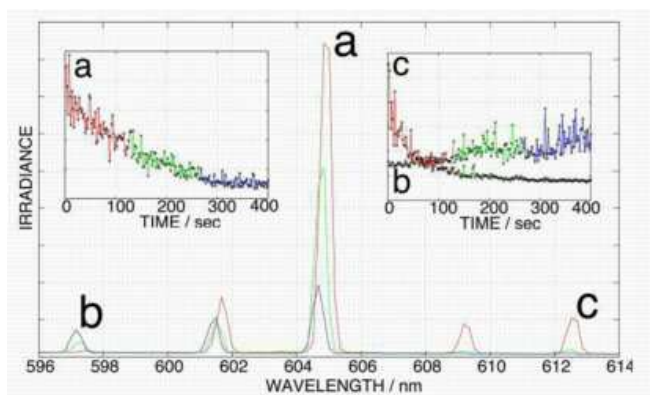}
 }
 \end{center}
\caption{Effect of photobleaching on the laser emission intensity:
short wavelength line b grows with progressive bleaching, while
longer wavelength lines a and c decrease.}
 \label{be50bleach}
\end{figure}

\medskip
On the other hand, the rhodamine~BE50-loaded samples under
investigation contained dye at a concentration of around one
Rh~BE50 molecule per 75 unit cells, and therefore bleaching caused
a detectable 4~nm shift of the fluorescence towards the blue
(bleach irradiance 0.5~MW/cm$^2$). Observing the laser emission
spectrum while bleaching the samples, a further consequence of the
blueshift was revealed: Blueshift of the fluorescence reduces the
overlap of the fluorescence band with the absorption spectrum, and
as a result laser modes at lower wavelengths will suffer less
losses with increasing bleaching. This is documented in
Fig.~\ref{be50bleach}, where the intensity of mode b with the
shortest oscillating wavelength increases, while
longer-wavelength-modes a and c decrease during the bleach
procedure. At the same time a 0.2~nm blueshift of the oscillation
wavelength was detected. We attribute this to a weak decrease of
the refractive index of the resonator material due to the smaller
polarizability of the dye debris. In contrast to the pyridine 2
samples, however, recovery of the fluorescence was not detected. As
the Rh~50BE molecules are considerably larger than pyridine 2
molecules their mobility in the molecular sieve framework is
severly hampered. So, diffusion of intact molecules into the
bleached volume occurs -- if ever -- on larger timescales than
minutes.

\section{Conclusion}
\label{conclu}

We have reported about a new form of microlasers based on
nanoporous molecular sieves containing embedded dyes. Size- and
shape-dependent laser properties were observed, in agreement with
theoretical predictions: Larg\-er hexagonal crystals ($WoF
\gtrapprox 10\,\mu$m) revealed multiline laser emission, while the
smaller ones ($WoF\approx 5\,\mu$m) oscillate on one single line.
The laser threshold power density of these small lasers is
approximately a factor of 3 lower than the threshold of $WoF
\approx 10\,\mu$m--lasers, and their differential efficiency is
almost an order of magnitude larger. Although these properties did
not depend on the type of embedded dye molecules, the
photostability seems to be affected by the size of the molecules.
Molecules which fit into the pores, such as pyridine 2, keep the
ability to diffuse in the pore framework of the molecular sieve.
Therefore photobleached molecules can be replaced by intact ones
diffusing into the luminescing volume.

\medskip
(Acknowledgement: This work was partially funded by the
\textit{Deutsche Forschungsgemeinschaft})

\end{document}